\documentclass[twocolumn,prd,aps,showpacs,superscriptaddress,nofootinbib]{revtex4}
\usepackage{amssymb}
\usepackage{epsf}
\usepackage{amsmath}
\usepackage{amsfonts}

\begin{document}

\title{The quadrupole moment of slowly rotating fluid balls}

\author{Michael Bradley\footnote{electronic address: michael.bradley@physics.umu.se}} 
\affiliation{Department of Physics, Ume\aa\ University, SE-901 87 Ume\aa, Sweden}

\author{Gyula Fodor\footnote{electronic address: gfodor@rmki.kfki.hu}} 
\affiliation{KFKI Research Institute for Particle and Nuclear Physics,
H-1525, Budapest 114, P.O.B. 49, Hungary}

\begin{abstract}

  In this paper we use the second order formalism of Hartle to study
  slowly and rigidly rotating stars with focus on the quadrupole
  moment of the object. The second order field equations for the
  interior fluid are solved numerically for different classes of
  possible equations of state and these solutions are then matched to
  a vacuum solution that includes the general asymptotically flat
  axisymmetric metric to second order, using the Darmois-Israel
  procedure. For these solutions we find that the quadrupole moment
  differs from that of the Kerr metric, as has also been found for
  some equations of state in other studies. Further we consider the
  post-Minkowskian limit analytically.  In the paper we also
  illustrate how the relativistic multipole moments can be calculated
  from a complex gravitational potential.

\end{abstract}

\pacs{04.40.Dg, 04.25.-g, 04.20.-q}

\maketitle

\section{Introduction}

Sources for the Kerr metric have been sought for long, but with little
success. The only known exact solution that can be matched to the Kerr
metric seems to be the rotating dust disk, with maximal allowed
angular momentum, found by Neugebauer and Meinel
\cite{NeugebauerMeinel}.  Some numerical studies with different
equations of state for the source find quadrupole moments differing
from that of the Kerr metric, see
e.g. \cite{Bertietal,HartleThorne,ChandrasekharMiller,LaarakkersPoisson}.
In some recent papers \cite{Cabezas,Martin} it is shown that in the
rigidly rotating case incompressible fluids and fluids with polytropic
equation of state cannot be sources of the Kerr metric in the
post-Minkowskian limit.  These results are in accordance with the
general expectation that the ellipsoidal shape of the rotating fluid
ball produces an extra contribution to the quadrupole moment which
should also be present in the corresponding quadrupole moment of the
external field \cite{Bertietal,Hernandez}.  For a general review of
relativistic rotating stars see \cite{Stergioulas}.

In the present work we use the second order formalism for slowly and
rigidly rotating stars, developed by Hartle \cite{Hartle}, to study
the quadrupole moment and its deviation from that of the Kerr metric
for different classes of possible equations of state.  For some
earlier applications of the formalism see, e.g.,
\cite{HartleThorne,ChandrasekharMiller,BEFR} and for a comparison with
numerical solutions of the full Einstein equations see
\cite{Bertietal}.
The relativistic multipole moments of the vacuum exterior metric up to
order two are calculated using an algorithm developed in \cite{FHP}.

The field equations for the fluid region to second order in the small rotational parameter
$\Omega$ will be solved numerically using fourth order Runge-Kutta.
When imposing an equation of state this system can be rewritten as
a first order system of ordinary differential equations for nine functions,
but a subsystem for six of these functions will be sufficient to study for
our purposes.
Assuming regularity at the centre the solutions will depend on three 
constants of integration, corresponding to zeroth order
central density or pressure, the magnitude of the angular velocity
and one more second order small constant. If the solutions are required
to be asymptotically flat this second order constant will be determined
in terms of the other constants. Due to scaling invariance in the angular
velocity it may be given a fixed value in the numerical runs. Hence, given
an equation of state, we need only vary the central pressure or density
when scanning the solution space.
The solutions are then matched to a second order axisymmetric
vacuum solution using the Darmois-Israel procedure \cite{Darmois,Israel,VeraMarsMacCallum}.  
This metric includes the general second order asymptotically
flat stationary axisymmetric vacuum solution as a special case. 

We also consider the post-Minkowskian limit analytically by expanding the
field equations in the small parameter $\lambda\equiv GM/r_1 c^2$ and
make a comparison with the results of \cite{Cabezas,Martin}.

The paper is organized as follows: In section \ref{Preliminaries} the
method is briefly described and the field equations are presented. 
Also the second order
vacuum metric is given and its relativistic multipole moments up to order two are calculated.  
The matching procedure is described in
section \ref{Matching} and the integration constants for the vacuum
solution are solved for in terms of the values of the interior
solution on the matching surface. In section \ref{Newvariables} the
equations are rewritten in a form suitable for numerical
integration. The results of the numerical runs are given in section
\ref{Numericalsolutions} and finally a post-Minkowskian analysis is
made in section \ref{post}.  

\section{Preliminaries}\label{Preliminaries}

To second order the metric of a slowly rotating axisymmetric object,
both in the interior fluid region and the outside vacuum region, can
be written as
\begin{eqnarray}\label{hartlemetric}
ds^{2}  &=&(1+2h)A^2dt^{2}-(1+2m)\frac{1}{B^2}dr^{2}-\nonumber\\
& &  (1+2k)r^{2}\left[  d\theta^{2}+\sin^{2}\theta\left(
d\varphi-\omega dt\right)  ^{2}\right]  \ , \label{ds}
\end{eqnarray}
where $\omega$ is first order and $h$, $m$ are $k$ are second order in
the rotational parameter \cite{Hartle}.  The requirements of
regularity at the centre and asymptotic flatness imply that the first
order function $\omega$ depends on $r$ only.  The second order
functions $h,m,k$ can be given as
\begin{eqnarray}\nonumber
h&=&h_0+h_2P_2(\cos\theta)\\\nonumber
m&=&m_0+m_2P_2(\cos\theta)\\
k&=&k_2P_2(\cos\theta)
\end{eqnarray}
where $h_0,m_0$ and $h_2,m_2,k_2$ are functions of $r$ only and
$P_2(x)=\frac{1}{2}(3x^2-1)$ is the second order Legendre polynomial.
This result follows from reflection symmetry in the equatorial plane,
from that the equations for $h$, $m$ and $k$ separate with the
ans\"atze $h=\sum_{i=0}^{\infty}h_i(r)P_i(\cos\theta)$ etc., and from
the fact that there are no inhomogeneous terms containing $\omega$ in
the equations for $h_i$, $k_i$ and $m_i$ for $i>2$. For more details
see \cite{Hartle}.

The matching of the two spacetime regions happens via the application
of a coordinate transformation $\varphi \rightarrow \varphi+\Omega t$
in the fluid region. It means that the inner fluid region rotates with respect
to the distant stationary observers with angular velocity $\Omega$.
This parameter $\Omega$ is considered to be the small expansion
parameter with respect to which $\omega$ is first order and the other
corrections $h,m,k$ are second order.
In addition to this, we can also rescale the
interior time coordinate first by a zeroth order constant $c_4$, 
and then later by a second order
small constant $c_3$, i.e. $t\rightarrow c_4(1+c_3)t$.  

The matter content of the interior is modelled by a perfect fluid
\begin{equation}
T_{ab}=(\rho+p)u_{a}u_{b}-pg_{ab} \ .
\end{equation}
The coordinate system used in (\ref{hartlemetric}) is assumed to be
comoving with the fluid, i.e. the 4-velocity is assumed to possess the
form
\begin{equation}
u^a=(1/\sqrt{g_{00}},0,0,0)=((1-h)/A,0,0,0)
\end{equation}
which also implies that the shear of the fluid is zero so it rotates
rigidly.

\subsection{The field equations}\label{sectionfieldeq}

In this subsection we list the field equations relevant to various orders. 
A similar system of equations with a slightly different choice of 
variables was given in \cite{Fodor}.
If no equation of state is specified then the
only equation one gets to zeroth order of the rotational parameter is
the pressure isotropy condition $G^1\,_1=G^2\,_2$ which reads as
\begin{equation}\label{G110}
B\frac{d^{2}A}{dr^{2}}
+\frac{d (rA)}{dr}\frac{d (B/r)}{dr}
+\frac{A}{r^2 B}=0
\ .
\end{equation}
Making use of $G^0\,_0=T^0\,_0$ and $G^1\,_1=T^1\,_1$ the energy
density and pressure of the non-rotating configuration reads as
\begin{eqnarray}\nonumber
\rho_0 &=&\frac{1}{r^{2}}\left[  1-\frac{d(rB^2)}{dr}\right]  \ ,\\
p_0 &=&\frac{1}{r^{2}}\left[  \frac{B^2}{A^2}\frac{d(rA^2)}{dr}-1\right]
\ .\label{pp}
\end{eqnarray}
To first order in the rotation parameter the only relation follows
from $G^3\,_0=0$
\begin{equation}\label{G30}
\frac{d}{dr}\left(  r^{4}\frac{B}{A}\frac{d\omega}{dr}\right)
+4r^{3}\omega\frac{d}{dr}\left(  \frac{B}{A}\right)  
=0\ . 
\end{equation}
The second order Einstein equations yield the following conditions.
From $G^1\,_2=0$ one gets 
\begin{equation}\label{G12}
r\frac{d}{dr}\left(  h_{2}+k_{2}\right)  
+r\left(  h_{2}-m_{2}\right)\frac{1}{A}
\frac{dA}{dr}-  h_{2}-m_{2}  =0\ . 
\end{equation}
The pressure isotropy condition in the angular directions,
$G^2\,_2=G^3\,_3$, gives
\begin{equation}\label{G22}
6\left(  h_{2}+m_{2}\right) -r^4 \frac{B^2}{A^2} 
\left( \frac{d\omega}{dr}\right)^2
+4r^3\omega^2\frac{B}{A}\frac{d}{dr}\left( \frac{B}{A}\right)=0 \ .
\end{equation}
The equality of the pressure in the angular and radial directions,
i.e.\ $G^1\,_1=G^2\,_2$ gives two equations.  After eliminating the
derivative of $h_2$ using (\ref{G12}) one obtains from the
$P_2(\cos\theta)$ part
\begin{eqnarray}\nonumber
&&2r\frac{B^2}{A}\frac{d A}{dr}\left(r\frac{d k_2}{dr}-m_2\right)
-2r^2Bh_2\frac{d }{dr}\left(\frac{B}{r}\right)
+ \\\label{G1122}
&&\ \ \ \ \ \ \ 
m_2-4k_2-5h_2-\frac{1}{3}r^4\frac{B^2}{A^2}\left( 
\frac{d\omega}{dr}\right)^2=0 \ ,
\end{eqnarray}
while the $\theta$-independent part gives an equation for $m_0$ and $h_0$.

The energy density function can be decomposed as $\rho=\rho_0+\rho_2$,
where $\rho_2=\rho_{20}+\rho_{22}P_2(\cos\theta)$ and $\rho_{20}$ and
$\rho_{22}$ are second order small functions of the coordinate $r$.
The analogous decomposition of the pressure is similarly given by
$p=p_0+p_2=p_0+p_{20}+p_{22}P_2(\cos\theta)$.
We get one more equation for $m_0$ and $h_0$ if we assume that the
equation of state $\rho=\rho(p)$ is unchanged to second order. For
details see \cite{BEFR}.

\subsection{Equations of state}

To complete the system one more equation, e.g. an equation of state
\begin{equation}\label{eos}
\rho=\rho(p) \, ,
\end{equation}
has to be specified. 
In this paper we will consider equations of state in the form:
\begin{equation}\label{eos2}
\rho=d_1 p + d_2 \left(\frac{p}{p_{c}}\right)^{1/\gamma}+d_3 \,\, .
\end{equation}
Here $p_c$ is the central pressure.  Equation (\ref{eos2}) includes
some of the more used approximations for the equation of state of
dense stars, like Newtonian polytropes, relativistic polytropes with
$d_1=1/(\gamma-1)$, linear ones as well as the incompressible case.

We use units such that the speed of light $c=1$ and $8\pi G=1$. If
then the unit of length is taken as $\alpha$ meter, the relation
between our units and SI units are given by
\begin{eqnarray}\nonumber
\rho_{SI}&=&5.358 \times 10^{25} \frac{\rho}{\alpha^2} \mathrm{kg/m^3} \\ \nonumber
x_{SI}&=&4.813\times 10^{42} \frac{x}{\alpha^2} \mathrm{kg m^{-1} s^{-2}} \\
d_{1SI}&=&d_1
\end{eqnarray}
where $x=\rho c^2, p, d_2$ or $d_3$.

Note that $m_0$ and $h_0$ do not appear in equations (\ref{G110}),
(\ref{G30}), (\ref{G12}), (\ref{G22}), (\ref{G1122}) and (\ref{eos})
and hence this subsystem for $A$, $B$, $\omega$, $m_2$, $k_2$ and
$h_2$ decouples. Notice also that the equations can be solved order by
order and that the equations contain $m_2$ only algebraically.  In
section \ref{Numericalsolutions} the system will be reformulated as a
coupled system of six first order ordinary differential equations.
Due to the requirement of a regular centre the solutions to this
subsystem will only depend on three constants of integration.

\subsection{Vacuum metric}\label{sectionvacuum}

In the exterior vacuum region we will use a frame adapted to the
asymptotically non-rotating observer.  Solving the field equations
detailed in Section \ref{sectionfieldeq} by imposing $p=\rho=0$ the
metric functions for the vacuum region are given as follows
\cite{Hartle, bfmp}\footnote{To make the norm of the timelike Killing field equal to unity at spatial infinity, 
the expression of $h_0$ in \cite{bfmp} has
been slightly modified through a second order coordinate transformation
of the time coordinate. }
\begin{displaymath}
A^2=B^2=1-2M/r
\end{displaymath}
\begin{displaymath}
\omega =\frac{2aM}{r^3}\ ,
\end{displaymath}
\begin{eqnarray}\nonumber
h_0 &=&-m_0=\frac 1{r-2M}\left( \frac{a^2M^2}{r^3}+c_2\right) \\\nonumber
h_2 &=&3c_1r\left( 2M-r\right) \log \left( 1-\frac{2M}r\right) +a^2\frac
M{r^4}\left( M+r\right)   \\\nonumber
&&+2c_1\frac Mr\left( 3r^2-6Mr-2M^2\right) 
\frac{r-M}{2M-r}\\\nonumber
&&+\left( 1-\frac{2M}r\right) 
r^2q_1  \label{eqh2} \\\nonumber
\end{eqnarray}
\begin{eqnarray}\nonumber
k_2 &=&3c_1(r^2-2M^2)\log \left( 1-\frac{2M}r\right) -a^2\frac M{r^4}(2M+r) 
 \\\nonumber
&&-2c_1\frac Mr(2M^2-3Mr-3r^2)+\left( 2M^2-r^2\right) q_1 \\
m_2 &=&6a^2\frac{M^2}{r^4}-h_2 \ .  \label{eqk2}
\end{eqnarray}
In this approximation, the slowly rotating solution is characterized
by the mass $M$, the first order small rotation parameter $a$, and the
second order small constants $c_1$, $c_2$ and $q_1$.  When $q_1$ takes
the value zero the metric is known to be the general asymptotically
flat stationary and axisymmetric vacuum metric to second order (see
e.g. \cite{bfp}). It can be easily checked that the solution is
of Petrov type D only if both $c_1$ and $q_1$ are zero. The metric is
then equivalent to the Kerr metric to second order with mass $M
\rightarrow M-c_2$.

When $q_1 \neq 0$ the metric cannot be asymptotically flat.  It is
important to keep in mind, however, that without the inclusion of this
constant the matching conditions on the zero pressure surface are
overdetermined in general \cite{MarsSenovilla,bfmp}.

\subsection{Multipole moments of the vacuum metric}

The notion of relativistic gravitational multipole moments for static asymptotically flat vacuum spacetimes
was developed in \cite{Geroch} by Geroch
and later extended to the stationary case in \cite{Hansen} and \cite{Thorne} by Hansen and Thorne. 
These moments are
defined on the 3-space of the timelike Killing trajectories.

A 3-space $( {\cal M}, h)$ with positive definite metric $h$ is said to be asymptotically flat if it can be
conformally mapped to a manifold $( { \cal \tilde M}, \tilde h)$ 
with the following properties

(i)  ${ \cal \tilde M}={\cal M}\cup\Lambda$, where $\Lambda$ is a single point,

(ii)  $\tilde\Omega\bigr|_{\Lambda}=\tilde\Omega_{,i}\bigr|_{\Lambda}=0$,  
$\tilde D_i \tilde D_j\tilde\Omega\bigr|_{\Lambda}=\tilde h_{ij}\bigr|_{\Lambda}$

\noindent
where $\tilde h_{ij}=\tilde\Omega^2 h_{ij}$.

From the timelike Killing vector field $K^a$ one constructs the two scalar functions $f=K^aK_a$ and $\psi$,
where the later is obtained from the curl of $K^a$, $\psi_{,a}=\epsilon_{abcd}K^b K^{c;d}$, that
is a gradient due to the vacuum equations. The complex gravitational potential then reads
\begin{equation}
\xi=\frac{1-{\cal E}}{1+{\cal E}}
\end{equation}
in terms of the Ernst potential ${\cal E}=f+i\psi$. It is given the conformal weight $-1/2$, so that
$\tilde \xi=\tilde\Omega^{-1/2}\xi$. For axisymmetric spacetimes the metric is completely determined by 
the value of the potential on the axis of symmetry \cite{Beig}.

The multipole tensors on the 3-space of timelike Killing trajectories, with coordinates $x^i$, $i=1,2,3$ 
and metric given by the projection operator $h_{ab}=-fg_{ab}+K_aK_b$,
are then defined recursively as
\begin{eqnarray}\nonumber
&&P^{(0)}(x^i)=\xi, \quad P^{(1)}_j(x^i)=\xi_{,j}, \\\nonumber
&&P^{(n+1)}_{k_1k_2...k_{n+1}}(x^i)=D_{<k_{n+1}}P^{(n)}_{k_1...k_n>}-\\
&&\frac{1}{2}n(2n-1)R_{<k_1k_2}P^{(n-1)}_{k_3...k_{n+1}>}
\end{eqnarray}
where $<k_1...k_{n+1}>$ denotes the symmetric and trace-free part. $D_i$ and $R_{ij}$ are the
covariant derivative and Ricci tensor, respectively, with respect to the 3-metric $h_{ij}$ \cite{Hansen}. 
The $\tilde P^{(n)}_{k_1...k_n}$ are defined correspondingly in terms of the tilded quantities.

For axisymmetric spacetimes the multipole moments are given entirely in terms of the scalar moments defined as
\begin{equation}
P_n=\frac{1}{n!}\tilde P^{(n)}_{k_1...k_n}n^{k_1}...n^{k_n}\bigr|_{\Lambda}
\end{equation}
in terms of the axis vector $n^i$ \cite{Hansen}.

In, e.g., \cite{FHP} and  \cite{Backdahl} algorithms for calculating the
multipole moments of a stationary axisymmetric spacetime
are developed. We here use the method of \cite{FHP} to calculate the moments up to
second order. First the metric is transformed to the canonical form
\begin{equation}
ds^2=f(dt-\tilde \omega d\varphi)^2-f^{-1}\left[e^{2\gamma}\left(d\rho^2+dz^2\right)+\rho^2d\varphi^2\right],
\end{equation}
where $f$, $\gamma$ and $\tilde\omega$ are functions of $\rho$ and $z$, through
the coordinate transformation
\begin{eqnarray}\nonumber\label{cordtrans}
\rho&=&Ar\sin\theta\left[1+h_0+h_2+k_2-\frac{3}{2}\sin^2\theta\left(h_2+k_2\right)\right]\\\nonumber
z&=&\left[(r-M)\left(1+2h_0-\frac{1}{2}\sin^2\theta\left(h_2+2k_2-m_2\right)\right) \right. \\
&&\left. +r^2A^2\left(h_{0,r}-\frac{1}{2}\sin^2\theta\left(h_{2,r}+k_{2,r}\right)\right)\right]\cos\theta\ 
\end{eqnarray}
that holds to second order.

From the timelike Killing vector $K^a=\delta^a_0$ we find the two potentials 
\begin{equation}
f=K^a K_a=g_{00}
=(1+2h)A^2-r^2\omega^2\sin^2\theta
\end{equation}
and 
\begin{equation}
\psi=-\frac{2aM}{r^2}\cos\theta, 
\end{equation}
obtained from the curl of $K^a$.

Introduce the coordinates $\bar \rho=\rho/(\rho^2+z^2)$,
$\bar z=z/(\rho^2+z^2)$. A suitable conformal factor is then $\tilde\Omega=\bar r^2\equiv \bar\rho^2+\bar z^2$ and hence $\tilde\xi=(1/\bar r)\xi$. As shown
in \cite{FHP}, the first (up to $n=3$) scalar moments $P_n$ are given by the
coefficients $m_n$ in the expansion
\begin{equation}
\tilde \xi (\bar\rho=0)=\Sigma_{n=0}^{\infty}m_n \bar z^n
\end{equation}
of $\tilde\xi$ on the axis, that in the original coordinates corresponds to $\theta=0$ (or $\pi$). 
Expansion of $\xi$ on the axis in the original coordinates gives to second order in the rotational
parameter
\begin{eqnarray}\nonumber\label{xiexp}
\xi&=&\frac{M}{r-M}+i\frac{Ma}{(r-M)^2}-\frac{M^2a^2}{r(r-M)^3}-\\
&&(h_0+h_2)A^2\left(\frac{r}{r-M}\right)^2.
\end{eqnarray}
From (\ref{cordtrans}) one finds that $r=z+M$ to zeroth order, which
is sufficient for the three last terms in (\ref{xiexp}) since they are
already first and second order. The first term need to be expanded to
second order in the rotational parameter, giving
\begin{eqnarray}\nonumber
\frac{M}{r-M}&=&\frac{M}{z-2h_0(r-M) -r^2A^2h_{0,r} }=\\&&\frac{M}{z}+
\frac{M}{z^2}\left[2h_0(r-M) +r^2A^2h_{0,r}\right] 
\end{eqnarray}
where once again $r=z+M$ can be used in all second order functions. Finally, expressing $\tilde\xi$ in $\bar z$,
one obtains to second order
\begin{equation}
\tilde \xi=M-c_2+iMa\bar z-M\left(a^2+\frac{16}{5}M^4c_1\right)\bar z^2 .
\end{equation}
Hence the mass is given by $M-c_2$, the angular momentum by $J=Ma$ and the quadrupole moment by
$Q=-M\left(a^2+\frac{16}{5}M^4c_1\right)$ . We will be interested in the relative deviation of the quadrupole 
moment  from that of the Kerr metric
\begin{equation}\label{deltaQ}
\frac{\Delta Q}{Q}\equiv\frac{Q-Q_{Kerr}}{Q_{Kerr}}=\frac{16M^4c_1}{5a^2}.
\end{equation}

An expansion of the exterior metric for large $r$ gives the
following leading terms of $g_{00}$ (with $q_1=0$)
\begin{equation}
g_{00}=1-\frac{2M\left(1-\frac{c_2}{M}\right)}{r}+
\frac{2MP_2(\cos\theta)\left(a^2+ \frac{16}{5}M^4c_1\right)}
{r^3}\ ,
\end{equation}
i.e., the associated Newtonian quadrupole moment reads as (cf.,
e.g. \cite{Landau})
\begin{equation}
Q_{11}=Q_{22}=-Q_{33}/2=2M\left(a^2+\frac{16}{5}M^4c_1\right)
\end{equation}
in an asymptotically Cartesian system with the 3-axis along the axis
of rotation. Hence it is, up to a factor of 4, the same as the relativistic moment.

\section{Matching}\label{Matching}

We here briefly describe the matching procedure, in which the
Darmois-Israel junction conditions \cite{Darmois,Israel} are used.
For more details the reader is referred to \cite{BEFR} and also to
\cite{VeraMarsMacCallum} for a general discussion on the matching of
axisymmetric bodies in second order perturbation theory.  In the fluid
region, the matching surface ${\cal S}$ is defined by the condition of
vanishing pressure, $p=0.$ In the limit of no rotation, the matching
surface is the 3-dimensional cylinder $r=r_{1}$.  For slow rotation
the equation of the matching surface ${\cal S}$ is
\begin{equation}\label{SW}
r=r_{1}+\xi 
\end{equation}
with 
\begin{equation}
\xi =-\left[\frac{p_{20}+p_{22}P_2(\cos \vartheta) 
}{\frac{d p_{0}}{d r}}\right]
_{\mid \;r=r_{1}}\equiv \xi_0+\xi_2 P_2(\cos\vartheta)
\end{equation}
where the constants
$\xi_0$ and $\xi_2$  are given by
\begin{eqnarray}\nonumber
\xi_0&=&
\frac{1}{12r B\frac{dA}{dr}\frac{d}{dr}\left(\frac{A}{B}\right)}\times
\label{xi0}\\
&&\left.\left[12r A^2\frac{dh_0}{dr}-12m_0\frac{d}{dr}\left(rA^2\right)
+r^4\left(\frac{d\omega}{dr}\right)^2\right]\right
\vert_{r=r_1}\nonumber
\end{eqnarray}
and
\begin{equation}\label{xi2}
\xi_2= -\left.\frac{\left(3A^2h_2+ r^2\omega^2\right)}
{3A\frac{d A}{d r}}\right\vert_{r=r_1} \, .
\end{equation}

In the vacuum exterior region suitable hypersurfaces for matching 
are determined by 
\begin{equation}\label{SV}
r=r_{1}+\chi \equiv \chi _{0}+\chi _{2}P_2(\cos \vartheta) ,
\end{equation}
where $\chi _{0}$ and $\chi _{2}$ are constants to be determined by the
matching conditions \cite{Roos}.

In order to find isometric embeddings of the matching surface ${\cal
  S}$ in the vacuum and fluid domains, we equate with each other the
respective induced metrics $ds^2_{ (v)}$, $ds^2$ and induced extrinsic
curvatures $K^{(v)}$, $K$ 
\begin{equation}\label{matchcd}
ds^2_{ (v)}\vert_{\cal S}=ds^2\vert_{\cal S}\qquad
K^{(v)}\vert_{\cal S}=K\vert_{\cal S}\ 
\end{equation}
where $K$ is defined by
\begin{equation}
K\equiv K_{ab}dx^adx^b\equiv h_a^{\,\,\, c}h_b^{\,\,\, d}n_{(c;d)}dx^adx^b
\end{equation}
in terms of the unit normal $n_a$ of the matching surface ${\cal S}$
and the projection operator $h_a^{\,\,\,b}=n_an^b+\delta_a^b$.
To adjust the different coordinate systems with each other we apply a
rigid rotation in the fluid region by setting
$\varphi\to\varphi+\Omega t$ where $\Omega$ is a constant.  Then we
re-scale the interior time coordinate $t\in R$ by $t\to c_4(1+c_3)t$
with further zeroth and second order constants $c_4$ and $c_3$ to be
determined from the matching conditions.

From the zeroth order matching conditions we get the following
relations (all functions here and in the following first and second
order relations are evaluated at the zeroth order matching surface
$r=r_1$):
\begin{equation}\label{eqMc4}
M=\frac{1}{2}r_1(1-B^2) \,\, , \,\, c_4=\frac{B}{A} \ ,
\end{equation}
\begin{equation}
r_1=\frac{A}{2B^2\frac{dA}{dr}}(1-B^2) \, .
\end{equation}
To first order we solve for $a$ and $\Omega$
\begin{equation}
a=\frac{Br_1^3}{3A(B^2-1)}\frac{d\omega}{dr}\,\, , 
\,\, \Omega=\frac{r_1}{3}\frac{d\omega}{dr}+\omega \, .
\end{equation}
From the second order equations we can solve for $c_1$,
$c_2$, $c_3$, $q_1$, $\chi_0$ and $\chi_2$ as
\begin{eqnarray}\nonumber
c_1=\frac{B^2}{9r_1^2A^2(B^2-1)^6}
\Bigl[r_1^4B^2(B^4-3)\left(\frac{d\omega}{dr}\right)^2\\
+36A^2h_2(1-B^4)+72A^2B^2(h_2+k_2)\Bigr] \label{eqc1}
\end{eqnarray}
\begin{equation}
c_2=\frac{\xi_0}{2}\left(B^2-1+2r_1B\frac{dB}{dr}\right)
-r_1B^2m_0 -
\frac{r_1^5}{36}\frac{B^2}{A^2}\left(\frac{d\omega}{dr}\right)^2
\end{equation}
\begin{equation}
c_3=\frac{r_1^4}{36A^2}\left(\frac{d\omega}{dr}\right)^2
+\frac{c_2}{r_1B^2}-h_0
\end{equation}
\begin{eqnarray}\nonumber
q_1&=&\frac{1}{9r_1^2A^2(B^2-1)^6}\Bigl\{18k_2A^2(B^4-1)(B^4-8B^2+1)
\\\nonumber
&& +216A^2B^2\ln B\left[2B^2(h_2+k_2)+  
 (1-B^4)h_2\right]+\\\nonumber 
&&36h_2A^2B^2(B^2-1)(B^4+B^2-8)+ \\\nonumber
&&r_1^4\left(\frac{d\omega}{dr}\right)^2B^2\left[(B^2-1)(2+11B^2-7B^4)+ \right.
\label{q1}\\
&& \left.
6B^2(B^4-3)\ln B \right]\Bigr\}\label{eqq1}
\end{eqnarray}
\begin{equation}
\chi_0=\xi_0\quad\hbox{and}\quad \chi_2=\xi_2\ ,
\end{equation}
where $\xi_0$ and $\xi_2$ are obtained from (\ref{xi0}) and
(\ref{xi2}).  Note that we did {\it{not}} impose the equation of state
(\ref{eos}) when calculating the matching conditions.  Hence an
appropriate matching can be done, i.e. the vacuum metric in section
\ref{sectionvacuum} is general enough for describing the exterior of
any axisymmetric rigidly rotating perfect fluid ball up to second
order.

\section{Equations and boundary values}\label{Newvariables}

In this section we provide a reformulation of the field equations
to a form more suitable for numerical integration.  By doing this we
can get higher precision at the origin where apparent singularities
arise, moreover the freely specifiable constants are identified more
easily this way.

\subsection{Integrating the zeroth order field equation}

In order to simplify (\ref{G110}) it is convenient to redefine the
functions $A$, $h$, $m$ and $k$ in terms of the function $\nu$ as
\begin{equation}
A=e^{\nu}, \quad h=\tilde h e^{-2\nu}, \quad m=\tilde m e^{-2\nu}, 
\quad k=\tilde k e^{-2\nu} .
\end{equation}
The equations simplify considerably due to the fact that only the
derivative of $\nu$ will appear. Hence we introduce the function $z$
by
\begin{equation}
\frac{z}{B}=r\frac{d\nu}{dr}+1 \, .
\end{equation}
Then the zeroth order equation (\ref{G110}) becomes first order in $z$
and algebraic in $B$ \cite{Fodorsph}
\begin{equation}\label{zeq}
B r \frac{dz}{dr}+2B^2+z^2-4Bz+1=0 \ ,
\end{equation}
furthermore, the pressure of the non-rotating configuration (\ref{pp})
takes the form
\begin{equation}
p_0=\frac{1}{r^2}\left(2Bz-B^2-1\right) \ .
\end{equation}

\subsection{Series expansion around a regular centre}\label{sec:series}

 For sufficiently regular configurations close to the centre the metric
coefficients can be given as power series in $r$.  Assuming that the
central pressure and density are finite it follows that $B(0)=z(0)=1$.
The assumption of smoothness of the configurations at the symmetry
centre, in the spacetime sense, implies that the odd coefficients in
the expansions of the basic variables are zero. 

Hence, assuming a smooth centre in the spacetime sense, the odd powers
will be omitted hereafter.  Plugging the expressions
\begin{eqnarray}\nonumber
B&=&1+b_1 r^2+b_2 r^4+...\\\nonumber
z&=&1+z_1 r^2+z_2 r^4+...\\\nonumber
\omega&=&\omega_0+\omega_1 r^2+\omega_2 r^4+...\\\nonumber
\tilde h_2&=&h_2^{(0)}+h_2^{(1)}r^2+h_2^{(2)}r^4...\\\nonumber
\tilde m_2&=&m_2^{(0)}+m_2^{(1)}r^2+m_2^{(2)}r^4...\\\label{expansion1}
\tilde k_2&=&k_2^{(0)}+k_2^{(1)}r^2+k_2^{(2)}r^4...
\end{eqnarray}
into the field equations together with the equation of state (\ref{eos}) justifies then that all coefficients can be
given in terms of $z_1$, $\omega_0$ and $h_1\equiv h_2^{(1)}$.

To zeroth order one obtains
\begin{equation}\label{serieh0}
h_2^{(0)}=m_2^{(0)}=k_2^{(0)}=0 \, ,
\end{equation}
then to second order
\begin{eqnarray}\label{serieh1}\nonumber
m_2^{(1)}&=&k_2^{(1)}=-h_2^{(1)}\equiv -h_1
\, , \\
\omega_1&=&\frac{2}{5}\omega_0(z_1-3b_1)\, ,\quad b_1=-\frac{1}{6}\rho_{\vert 2z_1}
\end{eqnarray}
and finally to fourth order
\begin{eqnarray}\nonumber
b_2&=&-\frac{b_1^2}{2}+\frac{1}{10}\left(z_1^2+3b_1^2-4b_1z_1\right)\frac{d \rho}{d p}\Bigr|^{}_{2z_1}
, \\\nonumber
z_2&=&b_1z_1-\frac{z_1^2}{2}-b_1^2\\\nonumber
h_2^{(2)}&=&\frac{\left(3z_1^2-16z_1 b_1 + 8 b_1^2 - 10 b_2\right)
\left(\omega_0^2 + 3h_1\right)}{42 \left(z_1 - b_1\right)} \\\nonumber
&&+\frac{\omega_0^2}{21}\left(17 b_1 - 5 z_1\right)
\\\nonumber
k_2^{(2)}&=&\frac{\omega_0^2}{6}(z_1-3b_1)
+\frac{h_1}{2}(b_1-z_1)-h_2^{(2)},\\\nonumber 
m_2^{(2)}&=&\frac{2\omega_0^2}{3}(z_1-3b_1)-h_2^{(2)}\\
\omega_2&=&\frac{\omega_0}{70}\left(z_1^2-36z_1b_1+74b_1^2-50b_2\right) . 
\end{eqnarray}

The central density and pressure are given by
\begin{equation}
\rho_{0c}=-6b_1\ ,\ \ \ p_{0c}=2z_1 \ .
\end{equation}
This shows that for realistic configurations $b_1<0$ and $z_1>0$,
consequently the $z_1-b_1$ term in the denominator of $h_2^{(2)}$ is
nonvanishing.

\subsection{System of differential equations}

Motivated by the results of the previous section it is advantageous to
define the new dependent variables $\beta$, $\zeta$, $\tilde \omega$,
$\hat\omega$, $\hat h$, $\hat k$ and $\hat m$ through
\begin{eqnarray}\nonumber
B&=&1+r\beta\ , \quad z=1+r\zeta \ , \\
\omega&=&\omega_0+r \tilde \omega\ , \quad \omega_{,r}=\tilde \omega + \hat \omega \ , \\\nonumber
\tilde h_2&=&r \hat h\ , \quad  
\tilde k_2=r(r^2\hat k-\hat h)\ , \quad
\tilde m_2=r(r^2\hat m-\hat h)\, .
\end{eqnarray}

The closed subsystem of equations (\ref{G110}), (\ref{G30}),
(\ref{G12}), (\ref{G22}), (\ref{G1122}) and (\ref{eos}) then gives six
first order differential equations for the quantities $\beta$,
$\zeta$, $\tilde \omega$, $\hat \omega$, $\hat h$ and $\hat k$, while
$\hat m$ can be solved for algebraically.
The equations for the zeroth order quantities are given by
\begin{eqnarray}
\frac{d\zeta}{dr}&=&-\frac{1}{rB}\left(-\zeta-3r\zeta\beta+r\zeta^2+2r\beta^2\right)\\
\frac{d\beta}{dr}&=&-\frac{1}{2rB}\left(r\rho_0+4\beta+3r\beta^2\right)
\end{eqnarray}
where the density $\rho_0$ is given by equation of state
$\rho_0=\rho(p_0)$ with the pressure given by
\begin{equation}
p_0=\frac{1}{r}\left(2\zeta+2r\zeta\beta-r\beta^2\right) \ .
\end{equation}
The equations for the first order quantities $\tilde\omega$ and
$\hat\omega$ are
\begin{eqnarray}
\frac{d\tilde\omega}{dr}&=&\frac{\hat\omega}{r}\\
\frac{d\hat\omega}{dr}&=&-\frac{1}{rB}\Bigl[r\left(4\omega_0+5r\tilde\omega+r\hat\omega
\right)\frac{d\beta}{dr}+4\left(2\beta-\zeta\right)\omega_0\nonumber\\
&&+\left(14r\beta+4-5r\zeta\right)\tilde\omega
+\left(7r\beta+5-r\zeta\right)\hat\omega\Bigr]
\end{eqnarray}
whereas the ones for the second order quantities $\hat h$ and $\hat k$
are long and hence are not given here. The corresponding homogeneous
equations, given by putting $\omega=0$ in (\ref{G12}), (\ref{G22}) and
(\ref{G1122}), are
\begin{eqnarray}\label{eqhhat}
\frac{d\hat h_{\mathrm{h}}}{dr}&=&\frac{2r\hat k_{\mathrm{h}}
-\beta\hat h_{\mathrm{h}}}{r(\beta-\zeta)B}
+\frac{\hat h_{\mathrm{h}}}{\beta-\zeta}\frac{d\beta}{dr}\\\label{eqkhat}
\frac{d\hat k_{\mathrm{h}}}{dr}&=&\frac{r\hat k_{\mathrm{h}}\left(2r\zeta-5r\beta
-3\right)+2\hat h_{\mathrm{h}}\left(\beta-\zeta\right)}{r^2B} .
\end{eqnarray}
These two last equations will be used when adjusting the parameters so
that the solutions become asymptotically flat.

Boundary conditions at $r=0$ are given as
\begin{eqnarray}\nonumber
\frac{d \beta}{d r}&=&b_1, \; \frac{d \zeta}{d r}=z_1, \; \frac{d \tilde \omega}{d r}=
\frac{d \hat \omega}{d r}=\frac{2}{5}\omega_0\left(z_1-3 b_1\right)
, \; 
\\
\frac{d \hat h}{d r}&=&h_1, 
\; \frac{d \hat k}{d r}=\frac{\omega_0^2}{6}\left(z_1-3b_1\right)
+\frac{h_1}{2}\left(b_1-z_1\right) \ .
\end{eqnarray}
The constants $b_1=-\rho_{0c}/6$ and $z_1=p_{0c}/2$ are related through the equation of
state (\ref{eos}), implying that we have three independent constants of integration (apart from
possible constants in the equation of state). One of this ($h_1$) will be determined through the
matching conditions on the zero pressure surface in terms of
the others when the requirement of asymptotic flatness is imposed.

There is a rescaling associated to the rescaling of the
rotational parameter $\omega_0$, following the rule $\omega_0
\rightarrow \gamma \omega_0$, which induces the transformation
\begin{eqnarray}\nonumber
&&\omega,\tilde\omega,\hat\omega\rightarrow\gamma\omega,\gamma\tilde\omega, 
\gamma\hat\omega \quad\quad \hat h,\hat m,\hat k
\rightarrow\gamma^2\hat h,\gamma^2\hat m,\gamma^2\hat k, \\\label{scale2}
&&\beta, \zeta, r \rightarrow \beta, \zeta, r  \ .
\end{eqnarray}
Due to this scale invariance of the equations the constant $\omega_0$
can be fixed.  All other configurations with a given equation of state
and specified central pressure can be obtained by rescaling.

The rescaling of the radial coordinate $r\rightarrow\alpha r$ induces
a change of the parameters in the equation of state, (\ref{eos2}),
given by
\begin{equation}\label{scaleeos}
d_1\rightarrow d_1, \quad
d_2,d_3,p_{0c}\rightarrow \frac{d_2}{\alpha^2},\frac{d_3}{\alpha^2},
\frac{p_{0c}}{\alpha^2} \ .
\end{equation}
The dependent variables then scale as
\begin{eqnarray}\label{scale1}
&&\beta,\zeta,\tilde\omega,\hat\omega,\hat m,\hat k
\rightarrow  \frac{\beta}{\alpha}, \frac{\zeta}{\alpha}, 
\frac{\tilde\omega}{\alpha},\frac{\hat\omega}{\alpha},
\frac{\hat m}{\alpha}, \frac{\hat k}{\alpha},\nonumber\\
&&\omega,\omega_0,\hat h \rightarrow \omega,\omega_0,\alpha\hat h \, .
\end{eqnarray} 
Using this scale invariance we can set one of the constants in the
equation of state to a fixed value. 

\subsection{Asymptotically flat solutions}\label{sec:asymp}

Asymptotically flat solutions can be obtained by considering suitable
linear combinations of homogeneous and particular solutions for $h_2$
and $k_2$, or equivalently for $\hat h$ and $\hat k$. 
These are solutions of the linear equations (\ref{eqhhat}) and (\ref{eqkhat})
and their corresponding inhomogeneous equations respectively.
Since we solve the equations order by order, and hence $\hat k$
and $\hat h$ do not appear in the lower order equations, we can add any
homogeneous solution to a given particular solution, i.e.,
$\hat k=\hat k_{p}+C\hat k_{h}$, $\hat h=\hat h_{p}+C\hat h_{h}$, where $C$ is an
arbitrary constant and the subscripts $p$ and $h$ refer to particular and
homogeneous solutions respectively. 
A solution to the field equations is asymptotically flat iff
$q_1=0$. Now the expression (\ref{eqq1}) is also linear in $k_2$ and
$h_2$, or equivalently in $\hat k$ and $\hat h$, it has a structure like
\begin{equation}
q_1=\alpha_1+\alpha_2 \hat k+\alpha_3 \hat h.
\end{equation}
Hence
\begin{eqnarray}\nonumber
q_1&=&\alpha_1+\alpha_2 (\hat k_{p}+C\hat k_{h})
+\alpha_3 (\hat h_{p}+C\hat h_{h})=\nonumber\\\nonumber
&&\alpha_1+\alpha_2\hat  k_{p}+\alpha_3 \hat h_{p}
+C(\alpha_2 \hat k_{h}+\alpha_3 \hat h_{h})\equiv\\
&&q_{1p}+Cq_{1h}
\end{eqnarray}
and $q_1=0$ can be obtained by choosing $C=-q_{1p}/q_{1h}$. Also the constant $c_1$ (\ref{eqc1}),
that is used to calculate the quadrupole moment, has a similar structure, and hence the 
resulting $c_1$ is given by $c_1=c_{1p}+Cc_{1h}$.

\section{Numerical solutions}\label{Numericalsolutions}

In this section we consider solutions with equations of state (\ref{eos2}),
with special attention to the quadrupole moments and their deviation
from that of the Kerr metric. In particular we determine the quantity $\Delta Q/Q$
defined in (\ref{deltaQ}).
The system was solved using fourth order Runge-Kutta.  In scanning the
parameter space, due to the scaling invariance,
without loss of generality, we fixed $\omega_0=0.1$ and the method for
obtaining asymptotically flat solutions,
described in  section \ref{sec:asymp} below, indirectly fixes the value of $h_1$. 
Hence, for a given equation of state, we only need to vary the central pressure,
$2z_1$. When considering different equations of state we use the scale
invariance (\ref{scaleeos}) and (\ref{scale1}) to fix one of the constants
$d_2$ or $d_3$. When performing the corresponding rescaling $r \rightarrow \alpha r$
the quantity $\Delta Q/Q$ is invariant.

The integrations were carried out
until the zero pressure surface was reached.  The accuracy of the code
was checked for the Wahlquist solution \cite{Wahl}.

\subsection{Incompressible fluids (interior Schwarzschild)}

\begin{figure} [!ht]
\epsfxsize=3.0in
\epsffile{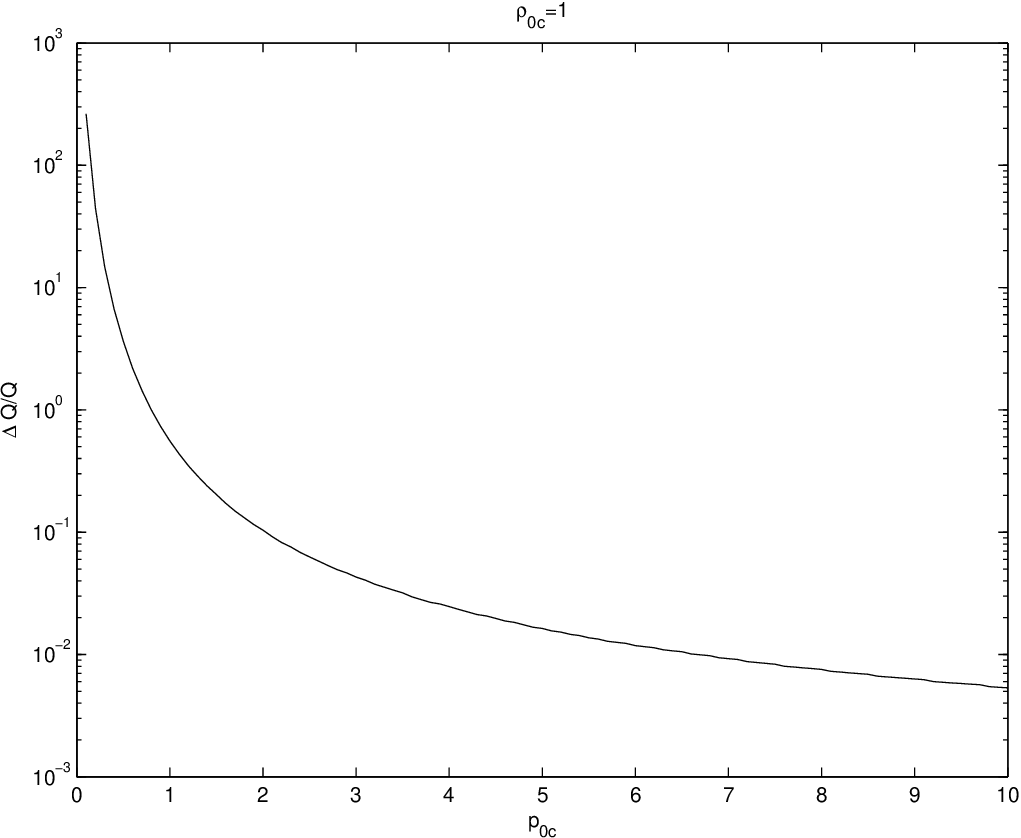} 
\vskip2mm
\caption{\small The quantity $\Delta Q/Q$ as function of central pressure for a sequence of
of fluid balls with constant density $\rho=1$.}
\label{fig:constantdensity}
\end{figure}

\begin{figure} [!ht]
\epsfxsize=3.5in
\epsffile{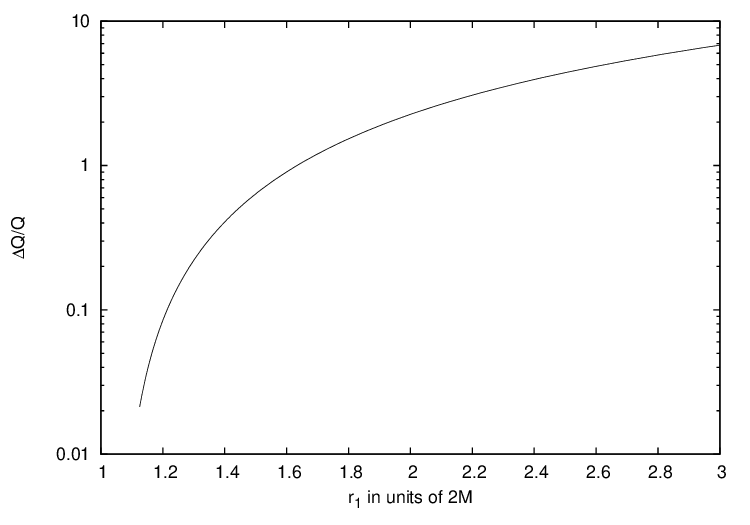} 
\vskip2mm
\caption{\small The quantity $\Delta Q/Q$ as function of radius $r_1/2M$ for a sequence of
of fluid balls with constant density $\rho=1$.}
\label{fig:constantdensity2}
\end{figure}

This case was already considered in \cite{ChandrasekharMiller}.
The numerically calculated quantity $\Delta Q/Q$, defined
in (\ref{deltaQ}),
as a function of the central pressure is depicted in Figure \ref{fig:constantdensity}.
In Figure \ref{fig:constantdensity2} the same quantity is given as
function of the radius of the corresponding non-rotating
configuration. When the central pressure approaches infinity and the radius the
Buchdahl limit $\frac{9M}{4}$, then $\Delta Q/Q$ approaches $0.0212$. Due to
the scaling invariance, (\ref{scaleeos}) and (\ref{scale1}), this limiting value is
independent of the density.

\subsection{Linear equations of state}

Linear equations of state 
\begin{equation}
\rho=d_1p+d_3
\end{equation}
are of interest not only because they could approximate some compact
objects, but also since in this class one might have a chance of
finding exact solutions, at least to zeroth order.

An exact spherically symmetric perfect fluid solution is given by the Whittaker metric \cite{Whitt},
with equation of state $\rho=-3p+\mu_0$. Its metric is
\begin{equation}\nonumber
ds^{2}=f_{0}dt^{2}-\frac{dr^2}{f_0\left(1-\kappa^2\mu_0r^2/2\right)}
-r^2\left( d\theta ^{2}+\sin ^{2}\theta d\varphi ^{2}\right) \, ,
\end{equation} 
where 
\begin{equation}
f_{0}=1+\frac{1}{\kappa ^{2}}\left[ 1-\arcsin\left(\kappa\sqrt{\frac{\mu_0}{2}}r\right)\sqrt{\frac{2}{\kappa^2\mu_0r^2}-1}\right]\, ,
\end{equation} 
and the pressure is
\begin{equation}
p=\frac{1}{2}\mu _{0}\left( 1-\kappa ^{2}f_{0}\right) .
\end{equation}
It is the non-rotating limit of the Wahlquist metric \cite{Wahl}, that
cannot be matched to an asymptotically flat vacuum region
\cite{bfmp,SarnobatHoenselaers}. However, there is another rotating
generalisation of Whittaker that can be matched, although its
closed-form is not known \cite{Mukumu}.

In the limit when the central pressure goes towards $\mu_0/2$,
corresponding to $\kappa\rightarrow 0$, the Whittaker metric
approaches the anti-de Sitter spacetime and the radius $r_1$ blows up
to infinity as $\sqrt{2/\mu_0}/\kappa$, but $r_1/2M$ goes as
$1+4\kappa^2/\pi^2$, i.e.  the surface approaches the event
horizon. This is not in conflict with the Buchdahl limit
\cite{Buchdahl}, since this spacetime does not meet the usual physical
requirements. Even if the these are weaker than in Buchdahl's original
work, see e.g. \cite{Andreasson}, the central density becomes negative
in the anti-de Sitter limit.  For the corresponding second order
rotating and asymptotically flat configuration we find that the
quadrupole moment approaches that of the Kerr metric in this limit.

Similar results hold for all configurations with $d_1<-1$. For all these $\rho_{0c}=-p_{0c}$
when $p_{0c}=d_3/(-d_1-1)$, and since the anti-de Sitter spacetime is a solution to our
zeroth order equations and the solution is uniquely given by central pressure and density
this is the resulting spacetime. Since the zero-pressure surface is further and further
pushed outward when approaching  the limit $p_{0c}=d_3/(-d_1-1)$, all these configurations
also become infinite in extension. The ratio $r_1/2M$ once again tends towards one, and we
also reobtain that the quadrupole moment approaches that for Kerr. Even if these spacetimes
are quite unphysical, a study of them still is helpful in understanding which conditions are
needed for a successful matching to the Kerr metric.

\begin{figure} [!ht]
\epsfxsize=3.3in
\epsffile{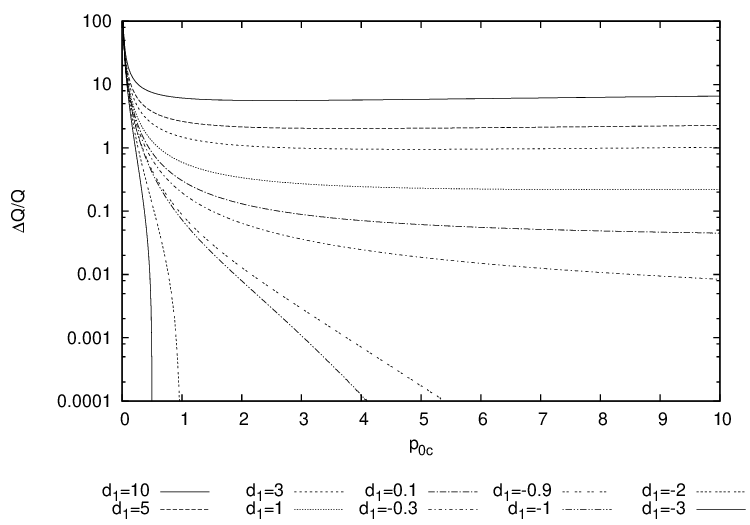} 
\vskip2mm
\caption{\small The quantity $\Delta Q/Q$ as a function of central pressure
for a sequence of fluid balls with linear equation of state $\rho=d_1 p+1$.}
\label{fig:linear}
\end{figure}

When performing the numerical runs, due to the scaling invariance,
(\ref{scaleeos}) and (\ref{scale1}), we put $d_3=1$. Negative values of $d_3$ are excluded
if the surface density should be larger than or equal to zero, and configurations with $d_3=0$
are not finite in size \cite{NilssonUggla}. The quantity $\Delta Q/Q$ for a sequence of values for
$d_1$ are given in Figure \ref{fig:linear}. As seen, for more realistic configurations with $d_1\geq1$ it
differs significantly from zero.

\subsection{Polytropes}

\subsubsection{Newtonian polytropes}

In a recent paper \cite{Martin} it was shown that slowly and rigidly
rotating polytropes cannot be sources of the Kerr metric in the 
post-Minkowskian limit. 
Here we numerically 
find that slowly and rigidly rotating polytropes with arbitrary strength of the 
gravitational field cannot be matched to Kerr either.

\begin{figure} [!ht]
\epsfxsize=3.3in
\epsffile{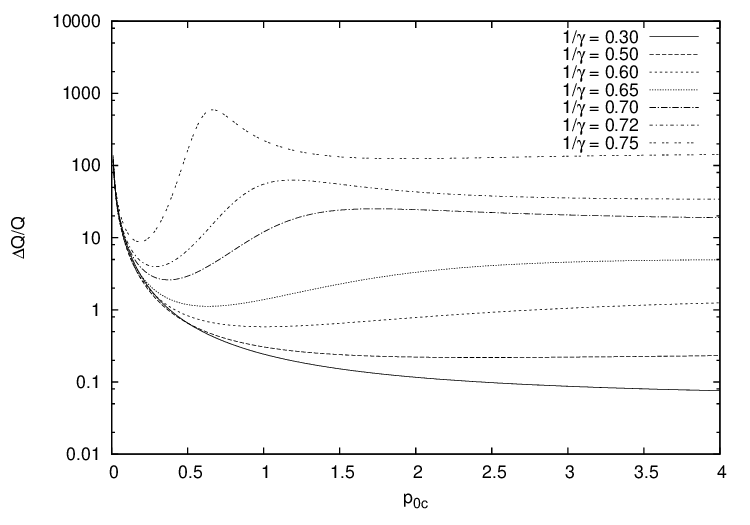} 
\vskip2mm
\caption{\small The quantity $\Delta Q/Q$ as a function of central pressure
for a sequence of Newtonian polytopes with equation of state $\rho=\left(\frac{p}{p_c}\right)^{1/\gamma}$.}
\label{fig:polytrope1}
\end{figure}

We use the equation of state
$\rho=d_2\left(\frac{p}{p_c}\right)^{1/\gamma}$ obtained by putting
$d_1=d_3=0$ in (\ref{eos2}). Due to the scaling invariance,
(\ref{scaleeos}) and (\ref{scale1}), we put
$d_2=1$.  In Figure \ref{fig:polytrope1} $\Delta Q/Q$ is shown as a
function of the central pressure for a sequence of  values of $\gamma$,
including the physically interesting cases $\gamma=4/3$ and $\gamma=5/3$. 
As shown in \cite{NilssonUggla} the configurations
are finite in size for $\gamma > 1.2295 $ ($1/\gamma <0.7695$) and infinite
for $\gamma<1.2$ ($1/\gamma>0.833$). In the interval in between the central
pressure determines whether the configuration is finite or not.
As seen $\Delta Q/Q$ never seems to approach zero. Furthermore, for physically reasonable configurations
with $p_c\leq\rho_c=1$, the quantity differs significantly from zero. 

\subsubsection{Relativistic polytropes}

\begin{figure} [!ht]
\epsfxsize=3.3in
\epsffile{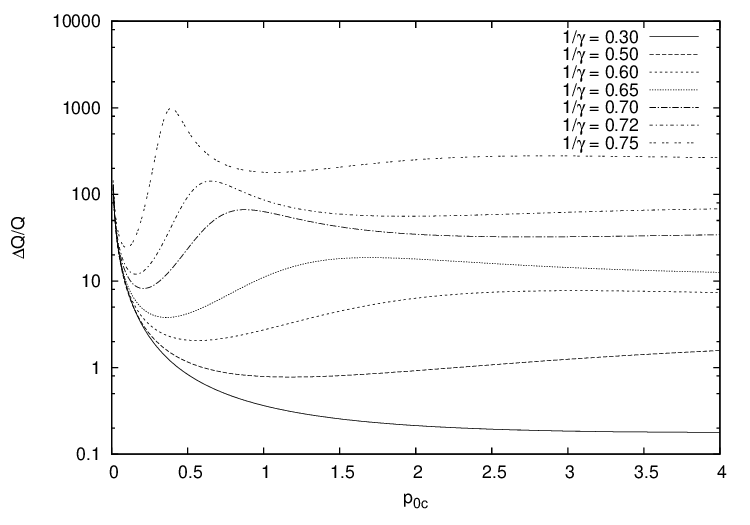} 
\vskip2mm
\caption{\small The quantity $\Delta Q/Q$ as a function of central pressure
for a sequence of relativistic polytropes with equation of state $\rho=\frac{1}{\gamma-1}p+\left(\frac{p}{p_c}\right)^{1/\gamma}$.}
\label{fig:polytrope2}
\end{figure}

For a fluid with one type of constituent particles, a relativistic polytrope is given by
$p=Cn^{\gamma}$, where $C$ is a constant, in terms of the particle density $n$
and the polytropic index $\gamma$. This equation is suitable, e.g., to describe an
ideal degenerate neutron gas.
Using the energy conservation equation for a perfect
fluid it is then easy to show that the equation of state becomes
\begin{equation}
\rho = \frac{1}{\gamma-1}p+d_2 p^{1/\gamma}
\end{equation}
in terms of the pressure $p$. As seen the equation of state approaches a linear
equation of state for large pressures and a Newtonian polytrope for low pressures. For 
a discussion of relativistic polytropes see. e.g., \cite{HeinzleRohrUggla}. 
The results of the numerical runs are given in Figure \ref{fig:polytrope2}. As expected the
results are similar to those of the Newtonian polytropes.

\subsection{Combined linear and polytropic}

When adding a constant to the Newtonian polytropes in Figure \ref{fig:polytrope1} typically
the local minima and maxima get less pronounced or disappear and for large central pressures 
the quadrupole moments get
closer to that of the incompressible case, but never smaller. For combined linear and polytropic
equations of state the curves become more similar to those of the relativistic polytropes in
Figure \ref{fig:polytrope2}, and when there is additative constant the limiting value of the
quadrupole moment for large pressures once again gets closer to that of the incompressible
case.

\section{The post-Minkowskian limit}\label{post}

Since we cannot solve the equations in the fluid region analytically,
we look at the weak gravity limit by making an expansion in the small
parameter $\lambda=M/r_1$, or in SI units $GM/r_1c^2$. Here $M$ is the
mass of the fluid ball and $r_1$ is its radius.  A similar study, using global
harmonic coordinates and the Lichnerowicz matching conditions \cite{Lichnerowicz}, 
was performed in \cite{Cabezas} and \cite{Martin}.

From an expansion of the equations in this parameter it turns out that
the pressure is one order higher in $\lambda$ than the density. This
can also be understood in a Newtonian context where the virial theorem
applied to a spherically symmetric configuration gives $\langle
p\rangle=k \lambda \langle \rho\rangle$, where $k$ is a constant of
order unity and $\langle p\rangle$ and $\langle \rho\rangle$ are the
average values of pressure and energy density. By considering the
balance between gravitational and centrifugal forces on the equator of
a rotating Newtonian fluid ball one obtains for the angular velocity
that $\omega<\sqrt{\lambda}/r_1$, so it seems reasonable to let
$\omega$ go as $\sqrt{\lambda}$. This is also consistent with that
$\omega$ or its derivatives only appear quadratically in (\ref{G22})
and (\ref{G1122}).

We first consider the post-Minkowskian expansion of the spherically
symmetric system. The functions $A$ and $B$ are expanded as
\begin{equation}
A=1+A_1+A_2+... \ , \quad B=1+B_1+B_2+... \ ,
\end{equation}
where subscripts refer to the order in $\lambda$. The freedom of
rescaling the time coordinate was used to put the zeroth order term in
$A$ to one. We do a similar expansion of the pressure $p=p_1+p_2+...$
and density $\rho=\rho_1+\rho_2+...$. Using the field equation
(\ref{G110}) we first obtain that $p_1$ is a constant. We put this
constant to zero so that a zero pressure
surface can be obtained. Then
\begin{equation}
B_1=-r\frac{dA_1}{dr}.
\end{equation}
The first nonvanishing terms of the density and pressure are
\begin{eqnarray}
\rho_1&=&\frac{2}{r^2}\frac{d}{dr}\left(r^2\frac{dA_1}{dr}\right)
\label{rho1}\\
p_2&=&p_c-\left(\frac{dA_1}{dr}\right)^2
-4\int\frac{1}{r}\left(\frac{dA_1}{dr}\right)^2dr .
\end{eqnarray}
To complete the system we have to impose an equation of state
$p=p(\rho)$, giving an equation for $A_1$. The differentiated version
of this equation reads as
\begin{equation}\label{a1eq}
\frac{dp(\rho_1)}{d\rho_1}\frac{d\rho_1}{dr}=-\rho_1\frac{dA_1}{dr},
\end{equation}
where $\rho_1$ is given by (\ref{rho1}).  We note that the total mass
$M$ of the configuration will be of the order $\lambda$, while the
radius $r_1$ will be of zeroth order.

The function $\omega$, which is first order in the rotational
parameter $\Omega$, has the post-Minkowskian expansion 
\begin{equation}
\omega=\omega_{1/2}+\omega_{3/2}+...,
\end{equation}
where the indices correspond to the appropriate fractional orders of
the quantities in $\lambda$. From the field equation (\ref{G30}) we
obtain that $\omega_{1/2}$ is a constant. This corresponds to a rigid
rotation of the system. It is necessary to keep $\omega_{1/2}$
nonzero, since this will provide the nonlinear source term in the
higher order equations.

The quantities second order in the rotational parameter, i.e. $h_2$,
$m_2$ and $k_2$, are expanded as
\begin{eqnarray}\nonumber
h_2&=&h_{21}+h_{22}+...,\\\nonumber
m_2&=&m_{21}+m_{22}+...,\\
k_2&=&k_{21}+k_{22}+...,
\end{eqnarray} 
where the second index at each new quantity indicate the order in the
post-Minkowskian parameter $\lambda$. To first order in $\lambda$
equations (\ref{G12}), (\ref{G22}) and (\ref{G1122}) yield 
$m_{21}=k_{21}=-h_{21}$ and
\begin{equation}
\frac{d^2h_{21}}{dr^2}+\frac{2}{r}\frac{dh_{21}}{dr}
-\frac{1}{2}\left(\frac{d\rho_1}{dA_1}+\frac{12}{r^2}\right)h_{21}
=\frac{\omega_{1/2}^2r^2}{6}\frac{d\rho_1}{dA_1} , \label{eqh21}
\end{equation}
where
\begin{equation}
\frac{d\rho_1}{dA_1}=2\frac{d^3A_1}{dr^3}\biggr/\frac{dA_1}{dr}
-\frac{4}{r^2}
+\frac{4}{r}\frac{d^2A_1}{dr^2}\biggr/\frac{dA_1}{dr} .
\end{equation}
We note that a particular solution to eq. (\ref{eqh21}) is given by 
$h_{21p}=-\omega_{1/2}^2r^2/3$.

Using the condition that $q_1$, given by Equation (\ref{q1}), should
vanish (to first order in $\lambda$) for an asymptotically flat solution and that the corresponding
$c_1$ is given by $c_1=c_{1p}-c_{1h}q_{1p}/q_{1h}$ as
described in Sec.~\ref{sec:asymp} one obtains
\begin{equation}\label{eqc1pm}
c_1=-\frac{5\omega_{1/2}^2}{48r_1^5
\left(\frac{dA_1}{dr}\right)\bigr|_{r=r_1}^5}
\frac{\frac{d}{dr}\log\left(
h_{21h}/\left(r^4\frac{dA_1}{dr}\right)\right)\bigr|_{r=r_1}}
{\frac{d}{dr}\log\left(
rh_{21h}/\frac{dA_1}{dr}\right)\bigr|_{r=r_1}} .
\end{equation}
Here $h_{21h}$ is a regular nonzero solution of the homogeneous version of 
(\ref{eqh21}). Note that in general $c_1$ goes to infinity as $1/\lambda^4$.

From the matching condition (\ref{eqMc4})
it follows that
\begin{equation}\label{eqMr12}
\frac{d A_1}{dr}\Bigr|_{r=r_1}=\frac{M}{r_1^2}
\end{equation}
where we assume that the mass $M$ is positive. 
In \cite{BEFR} it is shown that the rotating fluid ball is oblate in shape if
\begin{equation}\label{oblate}
k_{2\vert r=r_1}+\frac{\xi_2}{r_1}<0
\end{equation}
where $\xi_2$ is given by (\ref{xi2}). Substitution of
$k_{21}=-h_{21}=\omega_{1/2}^2r^2/3-h_{21h}$ in (\ref{oblate}) together with
(\ref{eqMr12})
gives that the fluid
ball is oblate iff
$h_{21h\vert r=r_1}>0$ to first order in $\lambda$.

For the incompressible case the equations can be integrated completely
and give for the asymptotically flat case
\begin{eqnarray}\nonumber
A_1&=&\frac{\rho_1r^2}{12}+a_0 \,, \quad B_1=-\frac{\rho_1r^2}{6} \,,
\quad \omega_{3/2}=\frac{\rho_1r^2\omega_{1/2}}{5} \,, \\ 
h_{21}&=&\frac{r^2\omega_{1/2}^2}{2} \,, \quad 
p_2=p_c-\frac{\rho_1^2r^2}{12} \,.
\end{eqnarray}
The integration constant $a_0$ can be transformed away by a first order (in $\lambda$) rescaling
of the time coordinate. The choice $a_0=-\rho_1r_1^2/4$ makes the
constant $c_4$ in (\ref{eqMc4}) equal to unity. 
Since $h_{21h\vert r=r_1}=5\omega_{1/2}^2r_1^2/6>0$ the ball is oblate.
From this solution the radius $r_1$ of the fluid ball and the parameters $M$ and $a$ of the exterior metric are given by
\begin{equation}
r_1=\frac{2\sqrt{3p_c}}{\rho_1} ,  \quad M=\frac{4}{3}\frac{(3p_c)^{3/2}}{\rho_1^2} , \quad
a=-\frac{24}{5}\frac{p_c \omega_{1/2}}{\rho_1^2}
\end{equation}
respectively.
The value of $c_1$ for asymptotically flat metrics is in this case
\begin{equation}
c_1=\frac{5}{1024}\left(\frac{\rho_1}{p_c}\right)^5\omega_{1/2}^2
\end{equation}
and the relative difference to the Kerr quadrupole moment is 
\begin{equation}
\frac{\Delta Q}{Q}=\frac{25\rho_1}{16p_c} 
\end{equation}
which diverges as
$1/\lambda$. By scaling the input parameters $\rho_{0c}$, $p_{0c}$ and $\omega_0$
in the numerical code used in the previous section as $\lambda$, $\lambda^2$ and 
$\lambda^{1/2}$ respectively, these values
are also obtained numerically in the limit $\lambda \rightarrow 0$.

For Newtonian polytropes, $p=p_c\left(\frac{\rho}{\rho_c}\right)^\gamma$,
where $\rho_c$ and $p_c$ are the central density and pressure respectively,
the integration of (\ref{a1eq}) gives the following equation for $A_1$ 
\begin{equation}\label{eqa1poly}
\rho=2\frac{d^2 A_1}{d r^2}+\frac{4}{r}\frac{d A_1}{d r}
=k(a_{1r_1}-A_1)^{\frac{1}{\gamma-1}}
\end{equation}
with $a_{1r_1}=A_1(r_1)$ being the value of $A_1$ at the zero pressure surface and 
\begin{equation}
k=\left(\frac{\rho_c^\gamma}{p_c}\frac{(\gamma-1)}{\gamma}\right)^\frac{1}{\gamma-1} .
\end{equation}
The homogeneous version of (\ref{eqh21}) then becomes
\begin{eqnarray}\label{eqh21h}
&&\frac{d^2h_{21h}}{dr^2}+\frac{2}{r}\frac{dh_{21h}}{dr}\\
&&\ \ \ -\frac{1}{2}\left(\frac{-k}{\gamma-1}
(a_{1r_1}-A_1)^{\frac{2-\gamma}{\gamma-1}} 
+\frac{12}{r^2}\right)h_{21h}=0 . \nonumber 
\end{eqnarray}

From (\ref{eqa1poly}) it follows that
\begin{equation}
\frac{d^2 A_1}{dr^2}\Big/\frac{d A_1}{dr}\Bigr|_{r=r_1}=-2/r_1
\end{equation} 
on the zero pressure surface $r=r_1$. Substitution of this and (\ref{eqMr12}) into (\ref{eqc1pm}) 
now gives
\begin{equation}\label{eqc12}
c_1=-\frac{5}{48}\frac{r^5\omega_{1/2}^2}{M^5}\frac{\left(\frac{d h_{21h}}{dr}\Big/h_{21h}-2/r\right)}{\left(\frac{d h_{21h}}{dr}\Big/h_{21h}+3/r\right)}\Biggr|_{r=r_1} .
\end{equation}
Let $h_{21h}\neq0$ be the solution of (\ref{eqh21h}) that makes the
spacetime asymptotically flat.  Substitution of
$h_{21}=-\omega_{1/2}^2r^2/3+h_{21h}$ in (\ref{eqq1}) then gives
\begin{equation}
q_1=\frac{1}{15 r^2}\left(9h_{21h}+3r\frac{dh_{21h}}{dr}-5\omega_{1/2}^2r^2\right)\Bigr|_{r=r_1}=0
\end{equation}
from which the denominator of (\ref{eqc12}) is
\begin{equation}\label{denom}
\left(\frac{d h_{21h}}{dr}\Big/h_{21h}+\frac{3}{r}\right)\Bigr|_{r=r_1}=\frac{5\omega_{1/2}^2}{3h_{21h}} .
\end{equation}
Note that $h_{21h}=0$ would give a non-zero $q_1$.

With the identifications $\gamma\equiv 1+1/n$,
\begin{equation}
A_1 \equiv -\lambda \Phi +a_{1r_1} \, , \,\,\, h_{21h}\equiv -\frac{\lambda}{2}\phi_2
\end{equation}
in terms of the functions $\Phi$ (that corresponds to the Newtonian potential for the non-rotating configuration)
and $\phi_2$ used in \cite{Martin},
and a rescaling of the $r$-coordinate $r\equiv \sqrt{2/\bar k}\, s$, with $\bar k\equiv k\lambda^{n-1}$,
equations (\ref{eqa1poly}) and (\ref{eqh21h}) read as 
\begin{equation}
\frac{d^2 \Phi}{d s^2}+\frac{2}{s}\frac{d \Phi}{d s}+\Phi^n =0
\end{equation}
and
\begin{equation}\label{eqphi2}
\frac{d^2 \phi_2}{d s^2}+\frac{2}{s}\frac{d \phi_2}{d s}+\left(n\Phi^{n-1}-\frac{6}{s^2}\right)\phi_2 =0
\end{equation}
respectively. Hence these are exactly the same as equations (23) and (41) in the paper \cite{Martin}
by Mart\'in et.al. By rescaling the time coordinate to first order in $\lambda$ the central data for their
function $\Phi$
can be met and hence we may rely on their theorem on page 10.
From their proof
the following inequality holds for regular solutions $\phi_2$ of (\ref{eqphi2})  
\begin{equation}
\frac{\phi_2^\prime(s_1)}{\phi_2(s_1)}-\frac{2}{s_1}<0 ,
\end{equation}
with $\phi_2^\prime\equiv d\phi_2/ds$ and $s_1=\sqrt{\bar k/2}r_1$, or
in our notation
\begin{equation}\label{inequality}
\frac{d h_{21h}/d r}{h_{21h}}\Bigr|_{r=r_1}-\frac{2}{r_1}<0 .
\end{equation}
That $\phi_2(s_1)\neq 0$ and hence
also $h_{21h}(r_1)\neq0$ is guaranteed by their lemma.
But (\ref{inequality}) is actually exactly one of the factors in the numerator of (\ref{eqc12}) and
hence we conclude that $c_1\neq 0$.

Substitution of (\ref{denom}) in (\ref{eqc12}) now gives that $c_1>0$ iff $h_{21h\vert r=r_1}>0$,
or equivalently if the fluid ball is oblate in shape. Hence $\Delta Q/Q=16M^4c_1/5a^2>0$ for
oblate Newtonian polytropes.

\section{Conclusions}

It seems that generically the modulus of the quadrupole moment for slowly and
rigidly rotating perfect fluid balls is larger than that of the Kerr metric. Also,
typically it becomes large for small central pressures.
For the incompressible case the ratio $\Delta Q/Q$ is a monotonically decreasing
function of central pressure. In the limit,
when central pressure goes to infinity, it approaches a small,
but positive value.  
Newtonian polytropes show a slightly more complicated behaviour, and $\Delta Q/Q$
possesses a minimum for $\gamma<2$. However, these minimal values are always larger
than $0.2$. The relativistic polytropes behave in a similar way, and the values of the
minima are even higher. For linear equations of state something interesting happens.
When the constant $d_1$ is less than $-1$, the configuration  becomes infinite in
extent for a finite central pressure. In this limit the zero pressure surface approaches
the event horizon and $\Delta Q/Q$ goes to zero. Also
the central regions of these fluids approach the anti-de Sitter spacetime. All of these 
solution are quite unphysical since they have
a central region with negative density. For the more realistic values of $d_1$, i.e. $d_1>1$,
the curves for $\Delta Q/Q$ show minima, but with values larger than $0.2$.

In a post-Minkowskian approximation the field equations simplify, but still
the equation of state can make the problem analytically unsolvable. 
Generically $\Delta Q/Q$ seems to go to infinity as $\lambda^{-1}$.
For the incompressible case the equations can be solved completely. and the result
agrees with \cite{Cabezas}. We further verify the result of  \cite{Martin} that Newtonian polytropes cannot
be matched to Kerr and also show that $\Delta Q/Q>0$ if the
rotating configuration is oblate.
The differences in approach were essentially that they used global harmonic coordinates
and the Lichnerowicz matching conditions, whereas we used the Darmois-Israel matching
procedure. The equivalence of the Lichnerowicz and Darmois-Israel matching conditions
has been demonstrated in the post-Minkowskian limit for the cases of Wahlquist and polytropes 
by Cuch\'i et. al., see \cite{Cuchi}. 

\acknowledgments

We thank P\'eter Forg\'acs for helpful discussions. MB gratefully
acknowledges the hospitality of the KFKI institute.  This work was
supported by OTKA-grants No. K61636 and NI68228.

\end{document}